\RequirePackage{lineno}
\documentclass[aps,prb,preprint,superscriptaddress]{revtex4}
\usepackage{graphicx}
\usepackage{bm}
\usepackage{amsmath}
\usepackage{amssymb}
\usepackage{hyperref}
\begin{document}

\title{Absence of Casimir regime in two-dimensional nanoribbon phonon conduction.}

\author{Zhao Wang}
\email{wzzhao@yahoo.fr}
\author{Natalio Mingo}
\affiliation{LITEN, CEA-Grenoble, 17 rue des Martyrs, 38054 Grenoble Cedex 9, France}

\begin{abstract}
In stark contrast with three-dimensional (3D) nanostructures, we show that boundary scattering in two-dimensional (2D) nanoribbons alone does not lead to a finite phonon mean free path. If combined with an intrinsic scattering mechanism, 2D boundary scattering does reduce the overall mean free path, however the latter does not scale proportionally to the ribbon width, unlike the well known Casimir regime occurring in 3D nanowires. We show that boundary scattering can be accounted for by a simple Mathiessen type approach for many different 3D nanowire cross sectional shapes, however this is not possible in the 2D nanoribbon case, where a complete solution of the Boltzmann transport equation is required. These facts have strong implications for the thermal conductivity of suspended nanostructures.
\end{abstract}

\maketitle
When the material size goes down to nanoscale, interfaces and sub-continuum phonon conduction phenomena play important roles in thermal transport. During the last few years there has been an increasing interest in the use of nanostructuring to control the thermal conduction in materials for various applications.\cite{Cahill2003,Balandin2008,Li2003,Li2003a} These experiments showed that the lowest lateral dimension (LLD) of nanostructures plays a key role in thermal transport. The first theoretical investigations of boundary scattering in thin wires were based on the works of Casimir,\cite{Casimir1938} Fuchs\cite{Fuchs1938}, Sondheimer,\cite{Sondheimer1950} Dingle,\cite{Dingle1955} and others.\cite{Ziman1960} These approaches allow us to interpret the results of recent suspended nanowire (NW) thermal transport measurements in terms of the boundary scattering limited mean free paths of phonons in the system.

It is known that the Mathiessen approach gives good description to the boundary scattering in cylindrical NWs.\cite{Dingle1955} However in fact the cross sectional shape of thin NWs may depart considerably from the cylindrical one. For example, the thermal conductivity of triangular and hexagonal shape InAs NWs has been recently measured, and its analysis requires to know how this shape affects boundary scattering.\cite{Lishipriv} Therefore, in the first part of this letter, we perform a systematic study of the effect of various shapes on boundary scattering. We conclude that the boundary and intrinsic scatterings can be considered to act additively, leading to a conveniently simple way to estimate mean free paths in the differently shaped NWs. We then investigate the validity of the Matthiessen rule for two-dimensional (2D) ribbons. We find that the Mathiessen approximation underestimates phonon conductance in nanoribbons.

The phonon mean free path $\lambda$ is the key parameter for modeling the thermal transport in nanostructured materials. The phonon mean free path in thin NWs is known to be proportional to the wire diameter, with a proportionality constant that depends on the wire's cross sectional shape. Accurate calculation of $\lambda$ is required for computing the thermal conductivity solving the Boltzmann transport equation (BTE) in the relaxation time approximation (RTA). Assuming that the boundaries are perfectly rough and that the wire length is much larger than the intrinsic mean free path $\lambda_{0}$ due to bulk scattering processes, and much larger than the LLD, $\lambda$ in a NW can be calculated using\cite{Ziman1960} 

\begin{equation}
\label{eq:1}
\lambda = \frac{ 3 \lambda_{0}}{4 \pi S} \int d\Omega \int  dS \cos^{2}{\chi} ( 1 - e^{{- l_{p}}/{\lambda_{0}}}),
\end{equation} 

\noindent where $l_{p}$ represents the phonon path length, $S$ is the cross section area, $\Omega$ is the solid angle, $\chi$ is the angle between the phonon path and the wire axis. This integral averages over all possible phonon orientations and positions in the cross section. 

\begin{figure}[ht]
\centerline{\includegraphics[width=13cm]{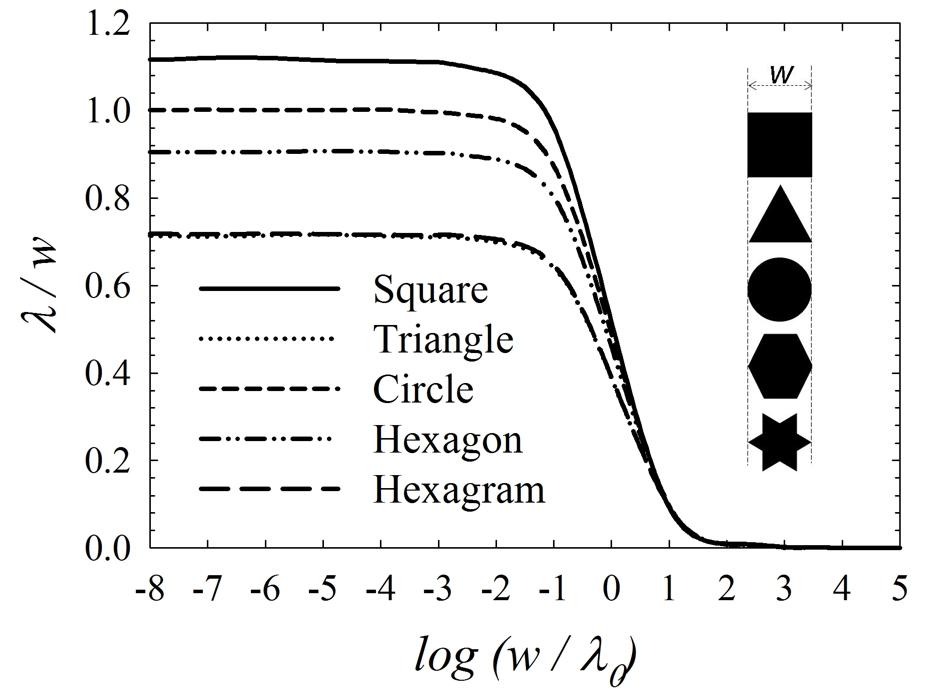}}
\caption{\label{fig:Fiveshapes}
$\lambda/w$ \textit{vs}. $w/\lambda_{0}$ for NWs in five different shapes (inset). $w$ is the cross section width.}
\end{figure}

\begin{figure}[ht]
\centerline{\includegraphics[width=13cm]{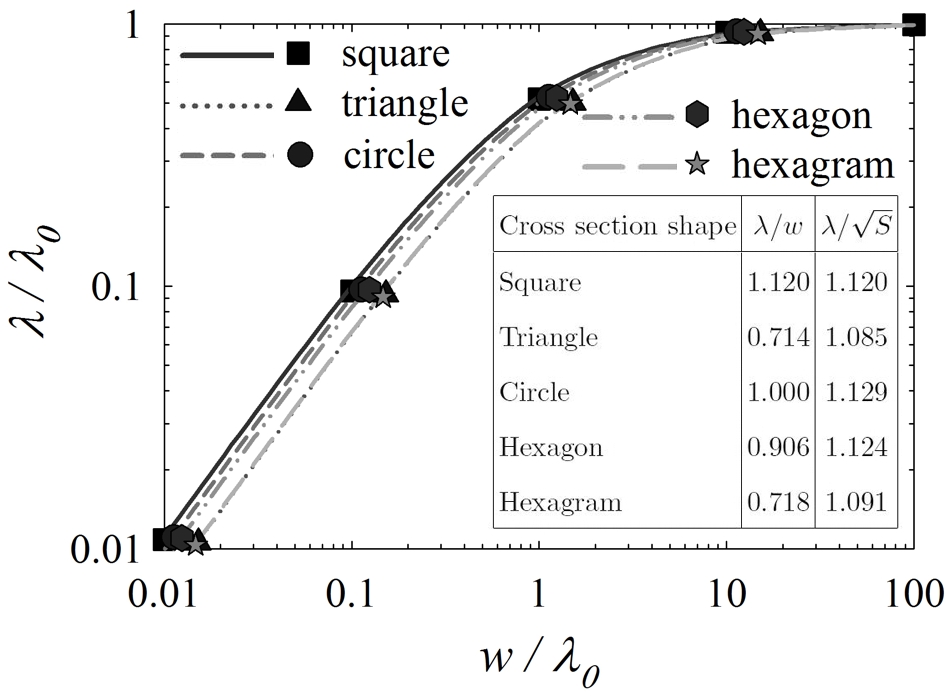}}
\caption{\label{fig:Casimir}
$\lambda/\lambda_{0}$ \textit{vs.} $w/\lambda_{0}$ in the non-linear region between ($\lambda_{0}>>w$) and ($w>>\lambda_{0}$). The symbols stand for results calculated by using Eq.\ref{eq:1} and the curves are plotted by using data computed from Eq.\ref{eq:3}. Inset: Values of $\lambda/w$ for a constant $S$, and those of $\lambda/\sqrt{S}$ for a constant $w$ ($\lambda_{0}>>w$).}
\end{figure}

To investigate the effect of different cross section, we use Eq.\ref{eq:1} to compute the $\lambda$ for NWs in five different cross section shapes: Square, triangle, circle, hexagon and hexagram. In Fig.\ref{fig:Fiveshapes}, it is shown that $\lambda/w$ saturates to a constant when $\lambda_{0}>>w$, where $w$ is the nanowire width defined in the same figure. The ratio $\lambda/w$ has previously been analytically obtained for the circular and square cross section cases, yielding $1.0$ and $1.12$ respectively.\cite{Ziman1960} The exact same values have been found in our calculation when $w/\lambda_{0}$ goes down to $10^{-6}$. The values for other shapes are given in the inset of Fig.\ref{fig:Casimir}. The linear relationship between $\lambda$ and $w$ holds approximatively until the NW diameter is approximately below one percent of $\lambda_{0}$. Beyond this limit, $\lambda$ starts to show dependence on $w/\lambda_{0}$. 

In the transition region between ($\lambda_{0}>>w$) and ($w>>\lambda_{0}$), we can compare the exact results from Eq.\ref{eq:1} with the one estimated by Mathiessen's rule,

\begin{equation}
\label{eq:3}
\lambda = \frac{1}{\frac{1}{Aw}+\frac{1}{\lambda_{0}}}
\end{equation}

\noindent where $A$ is the geometry specific constant tabulated in the inset of Fig.\ref{fig:Casimir} (first column). This empirical rule has been commonly used in the literature. Fig.\ref{fig:Casimir} shows this transition region by plotting $\lambda$ as a function of $w$, in which we compare our results computed by using Eq.\ref{eq:1} (symbols) with those calculated by using Eq.\ref{eq:3} (lines). It is shown that these two models give roughly the same estimation to $\lambda$ with slight difference. The mismatch between these two models is more significant for the triangle and hexagram cross sections, as expected since the edges of these two shapes are sharper than those of others.

When $w$ goes beyond the transition region to $w>>\lambda_{0}$, the value of $\lambda$ will be finally saturated to $\lambda_{0}$ when the size of the system reaches macroscopic bulk scale. For this we can see an inverse-linear trend in Fig.\ref{fig:Fiveshapes} when $w$ is roughly over ten times larger than $\lambda_{0}$ (Fig.\ref{fig:Casimir}). It can be seen that the saturation is slower for the cross section shapes with sharper edges. 

Boundary scattering affects phonon conduction quite differently in 2D nanoribbons than in three-dimensional (3D) nanostructures.\cite{Balandin2008} In narrow graphene ribbons, thermal conductivity can be significantly reduced by boundary scattering. In recent theoretical studies, different factors affecting the phonon transport in graphene were investigated by means of non-equilibrium\cite{Hu2009,Jiang2010} and equilibrium molecular dynamics (MD),\cite{Evans2010} and BTE approaches.\cite{Lindsay2010} The effects of graphene length and width, topological defect, uniaxial strain, edge roughness and hydrogen termination have been discussed. In particular, graphene boundary scattering effects were included in Ref.\onlinecite{Nika2009} by directly borrowing the 3D Casimir boundary scattering term. For a totally diffuse boundary, this would predict a strong boundary effect, with a mean free path that is proportional to, and of the same order as, the nanoribbon's thickness. This is justified for the case of graphene flake, investigated in Ref.\onlinecite{Nika2009}. However in this Letter, we show that the 2D long ribbons exhibit different boundary scattering dependence. If one only considers fully diffuse boundary scattering but no other scattering mechanism, the average phonon mean free path in graphene nanoribbons becomes infinity. In combination with an intrinsic scattering, a fully diffuse boundary will decrease the mean free path below that of the bulk. This effect is much weaker in 2D nanoribbons than in 3D NWs, e.g., 2D ribbons exhibit a slower dependence on thickness than the 3D nanostructures.

\begin{figure}[ht]
\centerline{\includegraphics[width=13cm]{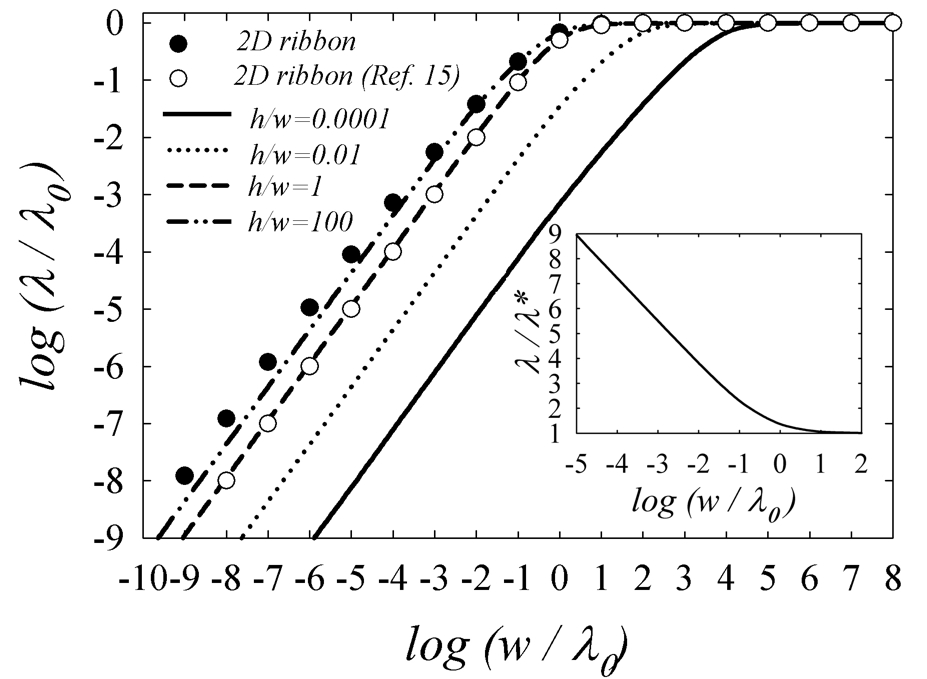}}
\caption{\label{fig:Graphene}
$\lambda/\lambda_{0}$ vs. $w/\lambda_{0}$ for rectangle rods with different aspect ratio $h/w$ (lines) and graphene ribbons with width $w$ (symbols). The solid circles stand for results of the present work and the empty ones stand for data calculated by using Eq.(19) in Ref.\onlinecite{Nika2009} ($\lambda^{*}$). Inset: The $\lambda$ of 2D ribbons divided by the one proposed by Ref.\onlinecite{Nika2009} $/\lambda^{*}$ vs. $w/\lambda_{0}$.}
\end{figure}

For a flat 2D ribbon, the $S$ in Eq.\ref{eq:1} becomes the ribbon width $w$ and the solid angle $\Omega$ becomes the angle $\theta$ between the phonon path and the ribbon edge. It yields

\begin{equation}
\label{eq:8}
\lambda_{2D} =\frac{2\lambda_{0}}{\pi w} \int_{0}^{w} dx \int_{-\frac{\pi}{2}}^{\frac{\pi}{2}} d\theta  \cos^{2}{\theta} (1-e^{\frac{-x}{\lambda_{0}\sin{\theta}}} ),
\end{equation}

\noindent where $x=l_{p} \sin{\theta}$. The integration in $x$ of Eq.\ref{eq:8} gives

\begin{equation}
\label{eq:2}
\lambda_{2D} =\frac{4\lambda_{0}}{\pi} \int_{0}^{\frac{\pi}{2}} d\theta  \cos^{2}{\theta} \left[ 1 - \frac{\lambda_{0}}{w} \sin{\theta} (1-e^{\frac{-w}{\lambda_{0}\sin{\theta}}})  \right].
\end{equation}

This expression is similar but not equivalent to the one for a thin film.\cite{Ziman1960} By plotting $\lambda_{\text{2D}}(w)/\lambda_{\text{3D-film}}(w)$ for a large range of $w$ (not shown), it is easy to verify that $0.85<\lambda_{\text{2D}}/\lambda_{\text{3D-film}}<1$, i.e. the thin film and 2D cases are very close, the largest deviation occurring for $w<<\lambda_0$. 

These equations also highlight another important qualitative difference with respect to the NW case: for 2D ribbons, the mean free path diverges in the absence of an intrinsic scattering mechanism. This is not the case for 3D NWs, in which boundary scattering results in a finite mean free path. Thus, it is not possible to derive a Casimir formula for 2D ribbons, and one cannot employ the approximated Mathiessen rule that we have earlier discussed for NWs. One of the correct ways to include the effect of boundary scattering in 2D nanoribbons is to perform the numerical integral in Eq.~\ref{eq:2}.
 
Fig.\ref{fig:Graphene} shows $\lambda/\lambda_{0}$ \textit{vs.} $w/\lambda_{0}$ for an infinitely-long rod with rectangular cross section $w \times h$, for different aspect ratios $h/w$. Like the trend of the square cross section shown in Fig.\ref{fig:Fiveshapes}, $\lambda$ is proportional to $w$ when $\lambda_{0}>>w$. In this region, the ratio $\lambda/w$ increases with increasing aspect ratio $h/w$. The value of $\lambda$ saturates to $\lambda_{0}$ at different values of $w$, depending on the aspect ratio, when the wire size becomes much larger than the intrinsic mean free path. It might have been naively expected that the $h<<w$ case would approach the 2D case. Instead, $\lambda$ in graphene roughly follows the trend of $\lambda$ in rectangular cross section NWs with high aspect ratio ($w<<h$). The reason is that, in a 3D thin film, $\lambda$ is strongly dominated by the phonon scattering on the top and bottom  surfaces when $h/w$ is small. However, this scattering mechanism does not exist in a 2D system like graphene. 

Finally, in Fig.\ref{fig:Graphene}, we compare our results for 2D ribbon phonon mean free path with those calculated using the expression proposed in Eq.(19) of Ref.\onlinecite{Nika2009}. The inset in Fig.\ref{fig:Graphene} clearly shows that the decrease of $\lambda$ for 2D ribbons is slower that that predicted by Eq.(19) in Ref.\onlinecite{Nika2009}. This slower decrease of $\lambda$ is in agreement with Green-Kubo MD simulations in Ref.~\onlinecite{Evans2010}, if $\lambda_{0}$ is considered to be in the order of several hundreds of nanometers\cite{Ghosh2008}. The thermal conductivities obtained by MD in Ref.~\onlinecite{Evans2010} are surprisingly large even for very narrow, rough edged ribbons. For example, for 10 nm wide ribbons, the simulation yields a thermal conductivity reduction of only about 50\% with respect to the infinite graphene case. The 3D formula of Ref.\onlinecite{Nika2009} would predict a much stronger reduction for this thickness, especially since the intrinsic mean free paths in graphene are known to be rather long. This is due to the fact that the phonon model used in Ref.\onlinecite{Nika2009} had originally been developed for 3D systems and it over-estimates the phonon scattering in graphene. This point is crucial for narrow graphene ribbons with width smaller than the graphene intrinsic mean free path (e.g. $\lambda_{0}\approx775$ nm as reported in experiments\cite{Ghosh2008}), e.g. for graphene with width in several tens of nanometers, the effect of phonon boundary scattering can be considerably weaker than previously thought.  

In conclusion, we have investigated the influence of sectional shape and dimensionality on the phonon mean free path of suspended NWs and nanoribbons. We have tabulated the Casimir geometric factors for five sectional shapes for the linear region (LLD $<< \lambda_{0}$). The exact dependence of $\lambda$ on $w$ has been shown to be reasonably well described by the Mathiessen approximation, using the tabulated factors. In contrast, we found that the effect of boundary scattering on the phonon mean free path of 2D ribbons is intrinsically different from that of 3D NWs. In particular, the 2D ribbon mean free path depends more slowly on system width, and it diverges in the absence of an intrinsic bulk scattering. As a result, neither a Casimir mean free path nor a Mathiessen rule can be defined for 2D ribbons. In the hypothetical case of vanishing intrinsic scattering, boundary scattering alone would lead to an infinitely long average mean free path in graphene, contrasting sharply with the thickness limited 3D case. This suggests that the phonon boundary scattering in graphene nanoribbons should be weaker than previously suggested.

We thank Arden Moore, Li Shi and D. A. Broido for helpful discussions. This work has been supported by the French ANR through the Accattone project, and by Fondation Nanosciences.

\end{document}